\documentclass[]{IEEEtran}
\usepackage{graphicx}
\usepackage[version=3]{mhchem}
\usepackage{amssymb}
\usepackage{color}
\usepackage{amsmath}
\usepackage{amsmath, multicol}
\usepackage[ruled,norelsize]{algorithm2e}

\DeclareMathOperator{\erf}{erf}
\newcommand{\Ugtrless}{%
	\mathrel{\kern0pt\mathop{\gtrless}\limits^{1}_{0}}%
}
\usepackage{textcomp}

\begin{document}

	\title{Performance Enhancement of Diffusion-based Molecular Communication with Photolysis}
	\author{\IEEEauthorblockN{Oussama Abderrahmane Dambri and Soumaya Cherkaoui}
			\\ 
		\IEEEauthorblockA{Department of Electrical and Computer Engineering, Universit\'e de Sherbrooke, Canada
			\\ 
			Email: \{abderrahmane.oussama.dambri, soumaya.cherkaoui\}@usherbrooke.ca}
	}
	\maketitle
	\begin{abstract}
		Inter-Symbol Interference (ISI) is the main challenge of the bio-inspired diffusion-based molecular communication. The degradation of the remaining molecules from a previous transmission is the solution that biological systems use to mitigate this ISI. While most prior work has proposed the use of enzymes to catalyze the molecules degradation, enzymes also degrade the molecules carrying the information, which drastically decreases the signal strength. In this paper, we propose the use of photolysis reactions, which use the light to instantly transform the emitted molecules so they no longer be recognized after their detection. The light will be emitted in an optimal time, allowing the receiver to detect as many molecules as possible, which increases both the signal strength and ISI mitigation. A lower bound expression on the expectation of the observed molecules number at the receiver is derived. Bit error probability expression is also formulated, and both expressions are validated with simulation results, which show a visible enhancement when using photolysis reactions. The performance of the proposed method is evaluated using Interference-to-Total-Received molecules metric (ITR) and the derived bit error probability. 
	\end{abstract}

	\section{Introduction}
The words of the great physicist Richard Feynman “There’s Plenty of Rooms at the Bottom” [1] were an inspiration for nanotechnology progress over the last century. The exponential growth of nano-devices’ use in every domain of science and industry paved the way to a big interest in nanonetworks design. Nanonetworks allow the communication between nano-devices at a nanoscale level. Nanonetworks can be used in several applications such as drug delivery and monitoring in the medical field. Several research works proposed using radio frequency communications to enable nanonetworks. However, using the traditional electromagnetic field at the nano level entails using the terahertz band with all its peculiarities and challenges. Indeed, THz suffer from a high path loss, which caused by molecular absorption [2]–[5]. 

One of the most promising solutions to communicate at such small level is using the method which nature adopted billions of years ago; utilizing molecules as information carriers between the transmitter and the receiver. The most basic method proposed for this bio-inspired paradigm is the Diffusion-based Molecular Communication (DbMC) [6], [7].

DbMC is based on the thermal fluctuations that allow molecules to move randomly in the medium using all possible degrees of freedom. This diffusive Brownian motion of molecules from the transmitter to the receiver is the wireless molecular communication used in the Biosystems, from bacterial colonies to the human brain [8]–[10]. Instead of using electromagnetic waves, the transmitter sends molecules as carriers of the information, which propagates in the medium and reaches the receiver. However, some molecules from a previous transmission usually stay in the medium because of their random walk. This causes an interference with the molecules of the next transmission, those introducing errors at the receiver and affecting the communication reliability.  To overcome this problem and mitigate the undesirable Inter-Symbol Interference (ISI), two categories of solutions have been proposed in literature; passive solutions [11]–[21] and active solutions [22]–[25].

Passive solutions can be categorized into; a) a group that simply ignores ISI [11] [12], b) a group that uses modulation-based methods [13]–[17], c) a group that uses pre-equalization methods [18] [19], and d) a group that optimizes the symbol time and the detection threshold [20] [21]. In [13], the authors proposed a new modulation method based on hybridization between Concentration Shift Keying (CSK) and Molecular Shift Keying (MoSK). They used two types of molecules and their corresponding concentrations to encode the information and mitigate ISI. The authors in [14] used the same principle, with the difference that instead of using molecular concentration shift, the information is coded in the difference between the two molecules concentrations, which allows a better ISI mitigation. The work in [15] proposes an ion protein based modulation, which can control the rate of molecules release to avoid ISI. The modulation method proposed in [16] uses a short block-length constrained graph to avoid ISI in a shorter delay than in [15]. However, this method can be used only in MoSK and not in CSK. Another newly proposed modulation technique uses the dynamic properties of calcium oscillation and propagation through coupled cells [17]. The information is encoded in the dynamic amplitude, dynamic period or both. Nonetheless, dynamic patterns in biological cells are very complex, and their robustness against the environment noise needs to be verified.

In [18], the authors use a pre-equalization method at the transmitter side to mitigate ISI, by means of two types of molecules; one for sending a primary signal and anther for sending a secondary signal. After applying the substitution operation at the receiver, the interference can be reduced. However, the receiver needs a delay to make this substitution operation after two consecutive emissions, which may decrease the achievable throughput. pre-equalization can also be applied at the receiver side as in [19], where two techniques are proposed; a linear equalization based on Minimum Mean-Square Error (MMSE), and a complex non-linear based on Detection-Feedback Equalizer (DFE) with better performance. Another passive solution is the optimization of the symbol time using a flow as in [20], where an optimization with drift is proposed to increases the speed of molecules and reduce ISI. In [21], the authors proposed an optimal delay for the receiver to shift its absorption interval and mitigate ISI. However, passive solutions such as those proposed in [13]–[21], either complicate the system or increase its delay, which decreases the achievable throughput.

Active solutions aim to physically remove the molecules from the medium and decrease ISI without increasing the delay. In [22] and [23], the authors proposed the use of neighboring receivers. The neighboring receivers compete with each other to absorb the molecules, which decreases ISI but also decreases the strength of the signal. Another active solution [24] uses enzymes to catalyze the degradation of molecules in the medium. Enzymes’ selectivity and recyclability allows transforming molecules so that they become unrecognizable by the receiver, which mitigates ISI. Instead of using the enzymes in all communication channel, the study in [25] proposed the deployment of a constant and limited number of enzymes around the transmitter or around the receiver. The study showed that deploying enzymes around the receiver with an optimal radius gives better results in mitigating ISI. Nevertheless, despite their advantages, the previously discussed active solutions decrease the signal strength [21]–[25]. The neighboring receivers absorb the molecules carrying the information, and the enzyme degrades not only the remaining molecules in the medium, but also the molecules carrying the information.

To mitigate ISI without decreasing the signal strength or increasing the delay, we need faster reaction to catalyze the remaining molecules in the medium and a catalyzer that can be switched ON/OFF. Using light can help satisfying these two criteria. The photolysis is a chemical process where chemical bonds are broken as the result of a light energy transfer [26]. Photolysis occurs with a very fast, first order degradation rate. The rate depends upon several chemical and environmental factors such as the light adsorption properties of the medium, the intensity of light radiation and the reactivity of the target chemical. The most famous photolysis reaction is the photosynthesis in plants, where light energy splits water molecules and produces oxygen. However, the most studied photolysis reaction is the one that triggers the daily vitamin D formation in human skin after sunlight exposure.

In our previous work [27], we proposed the use of UV light to instantly degrade provitamin D3 molecules into previtamin D3 as shown in Figure 1. The UV light is turned ON at an optimal time, i.e. when the maximum number of molecules is absorbed at the receiver, to degrade only the remaining molecules and mitigate ISI without decreasing the strength of the signal. These kind of photolysis reaction happens daily in human skin and it could be used safely in medical applications inside the body.

	\begin{figure}
	\centering
	\includegraphics[width=\linewidth]{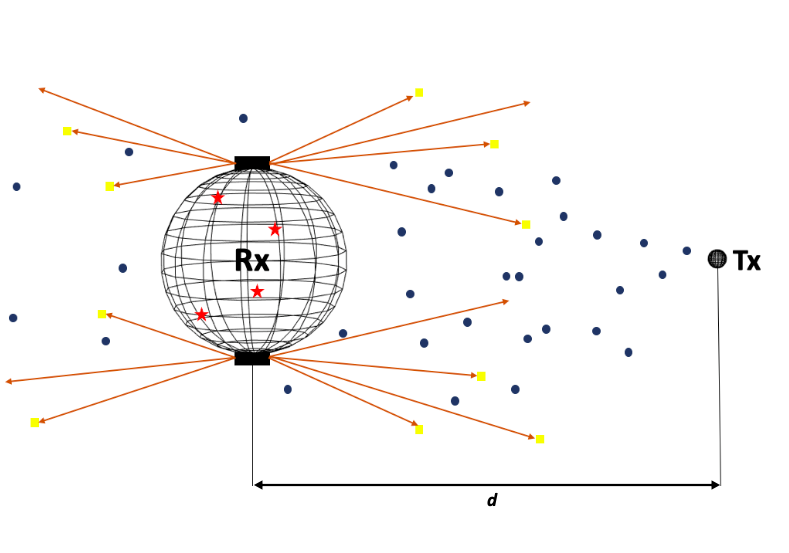}
	\caption{System model showing molecules propagation (shown as blue circles), released from the transmitter (Tx) to the 3D spherical receiver (Rx) away with distance $d$. The received molecules (shown as red stars) are inside the 3D receiver Rx. At time $T_{op}$, the light emitters (shown as black rectangles) up and down the receiver Rx emit light waves (shown as green arrows) hitting the molecules, which they absorb the light energy and instantly transform to photolysis reaction's products (shown as yellow squares).}
	\label{fig:1}
    \end{figure}

This paper is an extension of the work proposed in [27], where we introduced the system concept and some preliminary simulation results. In this paper, we propose an analytical model for analyzing diffusion-based molecular communication systems in general, when they are exposed to any light in a larger spectrum than solely UV. The main contributions of the paper are summarized as follows: 

\begin{enumerate}
\item	We present a lower bound expression on the expected number of molecules observed at a passive receiver when using a binary modulation and a light beam as a molecule catalyzer.

\item	We present an analytical comparison between a scenario without reaction, a scenario with enzymatic reaction and the proposed photolysis reaction scenario. 

\item	We derive the expression of the optimal time a which to turn the light ON, giving more chance to as many molecules as possible, to reach the receiver before triggering the photolysis reaction.
   
\item	We derive the bit error probability as a function with a threshold value to evaluate the proposed analytical model and the simulation results.

\end{enumerate}

The rest of the paper is organized as follows. In section II, we present the proposed system model that comprises a single transmitter using binary CSK modulation, and a single receiver, which can emit a controlled flash light. In section III, we present the derived lower bound expression on the expected number of molecules at the receiver when exposing the medium to light. A comparison between a scenario without reaction, a scenario with enzymatic reaction and the proposed photolysis reaction scenario is presented. The expression of the optimal time at which to turn the light ON and the bit error probability as a function with a threshold are derived in the section IV. In section V, we discuss the obtained results. Conclusion and future work are described in section VI.

	\section{System Model}
The proposed system model in this paper is similar to the system studied in [27], where we consider that the DbMC process takes place inside a 3D fluidic environment. The system contains a fixed point transmitter, a spherical receiver with a radius $r$ and a communication channel with a distance $d$ as shown in Fig. 1. The modulation of the signal at the transmitter in DbMC can be done by shifting the concentration of molecules, changing their types or shifting their time of release. In this study, the signal is modulated with a binary Concentration Shift Keying (CSK). The transmitter sends an impulse of molecules to represent bit “1” and to reduce the energy consumption, it sends nothing to represent bit “0”. The receiver is a 3D spherical passive observer, which can only count the number of molecules while they diffuse through it and cannot interact with them or change their behavior. If we neglect the interaction among the independent propagating molecules, and if assume that all molecules have an identical spherical shape and the same diffusion constant $D$, we can deduce that their diffusion follows a continuous-time stochastic process. This Brownian motion that allows molecules to diffuse towards the receiver can be described with Fick's second law as follows:

\begin{equation}
\frac{\partial S}{\partial t} = D \nabla^2 S,
\label{eq:1}
\end{equation} Where $S$ is the concentration of the released molecules, $t$ is the time and $D$ is the diffusion coefficient that can be calculated with the following equation:

\begin{equation}
D = \frac{k_B T}{6 \pi \eta R},
\label{eq:2}
\end{equation} Where $R$ is the radius of the molecules, $T$ is the temperature in kelvin, $k_B$ is the Boltzmann constant, and $\eta$ is the viscosity of the medium ($\eta$ $\approx$ $10^{-3} kg.m^{-1}s^{-1}$ at temperature 25°C) [24].

We assume the existence of a light nano-emitter that can emit a monochromatic light with a predetermined wavelength $\lambda$ and in a controlled optimal time $T_{op}$. The light nano-emitter will wait for a maximum number of molecules to reach the receiver and then, it switch ON the light emission to degrade the remaining molecules in the medium. To implement the nano-emitter on the top and the bottom of the receiver as shown in Fig. 1 and use it safely inside the human body, we can take advantage of the specific and selective receptors that we find embedded in cell membranes. We can also use the DNA hybridization to safely bind the nano-emitter with the receiver as proposed in [28]. By inserting a sequence of DNA on the surface of the receiver, and its complementary on the surface of the nano-emitter, they bind together with hydrogen bonds when they get close to each other, which binds the nano-emitter to the receiver. The implementation of the nano-emitter will be noninvasive and safe because of its tiny size. While the miniaturization of monochromatic light generators is reaching the millimeter level as the photochemical UV-LED proposed in [29], with the fast progress of nanotechnology, reaching the micro and nano levels is just a matter of time.

The light in this proposed system is the reaction catalyzer, which instantly transform the molecules that absorb the light energy by breaking specific chemical bonds. Therefore, the transforming molecules will no longer be recognized by the receiver and they cannot cause ISI. The reaction that breaks chemical bonds by using the absorbed light energy is called a photolysis reaction.

	\subsection{Photolysis Reaction}
The photodissociation famously known as the photolysis is a reaction where a chemical bond of a chemical compound is broken down by a photon [26]. This photon must have a sufficient energy that allows the bond to overcome its dissociation barrier. The photons in the infrared spectral range do not have enough energy for a direct photodissociation of molecules, because the photon’s energy is inversely proportional to its wavelength. Thus, the energy of the visible light or higher electromagnetic waves as UV light, X-rays and gamma rays are usually the ones involved in this reaction. The photolysis has been intensively studied in different domains such as in the atmosphere reactions that creates the ozone layers [30]. It is also studied in photosynthesis as part of the light-dependent reactions [31], and in astrophysics, where photolysis plays an important role in interstellar clouds formation [32]. However, even if the wavelength, the chemical compound and its by-product change in every study, the photolysis principle stays the same. Each chemical compound needs a specific amount of energy to break a bond, and when the electromagnetic wave passes through it with the specific wavelength, it transfers the energy of the wave to the bond and break it. In the case of the ozone layer, the energy needed to split oxygen molecules into two individual atoms in the Mesosphere can be provided with an ultraviolet photon with a wavelength between 200 and 300 nm [30]. The by-product of this photolysis reaction combines with an unbroken $O_2$ to create the ozone $O_3$ in the Stratosphere. On the other hand, the light-dependent
 reactions in the photosynthesis need the visible light, to split water molecules, especially the blue with a wavelength between 450 - 495 nm and the red with a wavelength between 620 - 750 nm [31]. In human body, we can find photolysis in the skin, where the provitamin $D_3$ transforms to previtamin $D_3$ when exposed to the UV photons of the sun light [33]. Then, the previtamin $D_3$ transforms to vitamin $D_3$ with a thermal isomerization as in the reaction shown in Fig. 2 [34].
 
	\begin{figure}
	\centering
	\includegraphics[width=\linewidth]{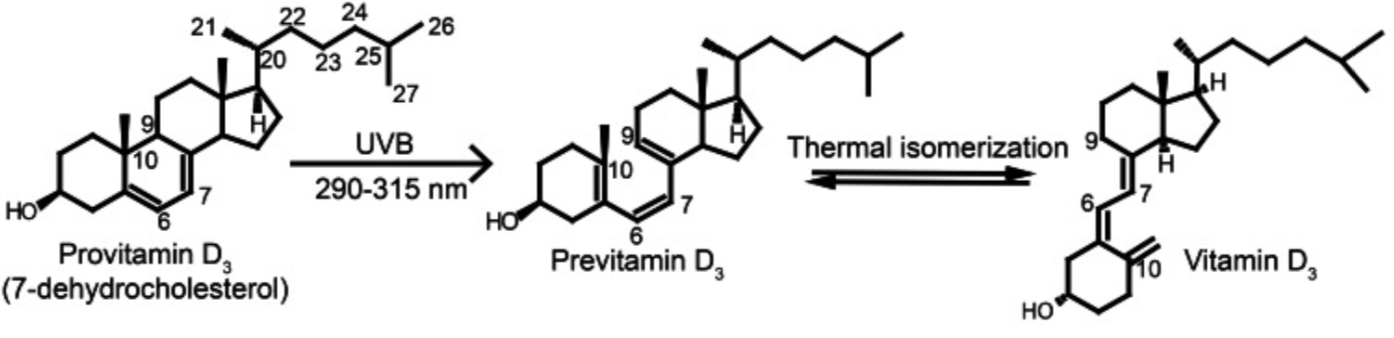}
	\caption{Vitamin $D_3$ Synthesis with a photolysis  and a thermal reactions. [34]}
	\label{fig:2}
    \end{figure}

In our previous study [27], we proposed the use of provitamin $D_3$ molecules as carriers of the information and the UV light as a catalyzer to the photolysis reaction in order to mitigate ISI. We chose 298 nm wavelength for the UV light photons because it is the wavelength where the absorption of provitamin $D_3$ is in its maximum as proved in the study [35]. In this paper, we generalize the use of the catalyzer’s wavelength from 298 nm to a spectrum range between [200 - 750 nm]. We can use photons with any wavelength inside that spectrum, with a condition that the chosen wavelength should be absorbed by the molecules carrying the information to dissociate them. 

		\begin{figure}
		\centering
		\includegraphics[width=0.7\linewidth]{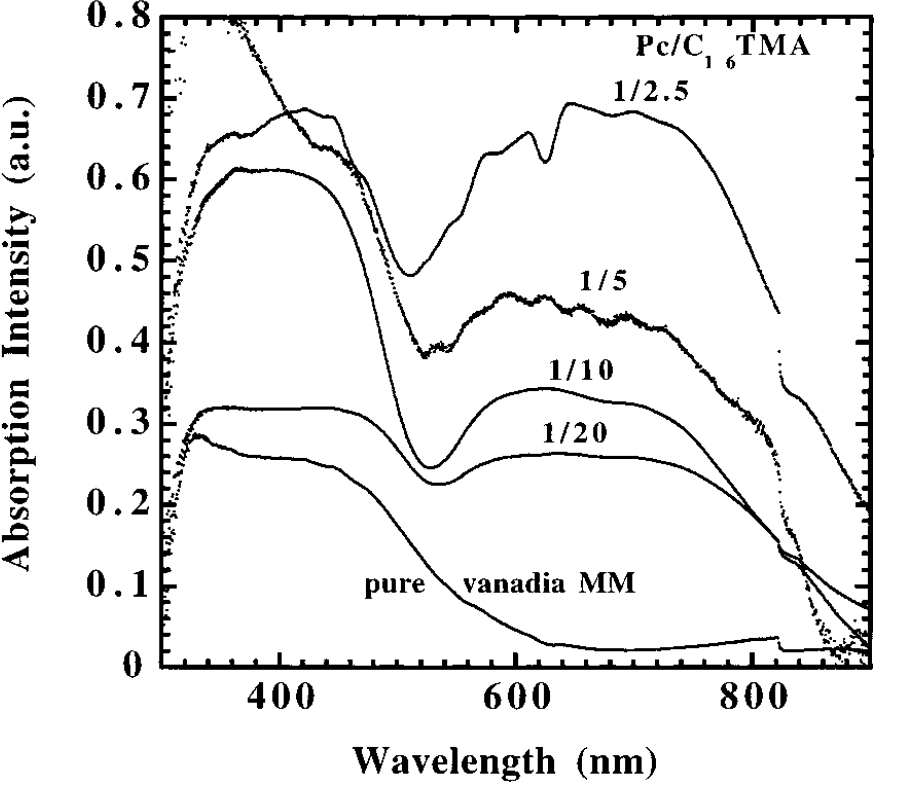}
		\caption{Optical absorption spectra for the pure and the doped vanadia. [36]}
		\label{fig:3}
    	\end{figure} 
    
The choice of molecules carrying the information which are sensitive to the chosen wavelength can be either from examples already exists in nature as the chromophores and provitamin $D_3$  or from bioengineered molecules. The chromophores are sensitive to visible light, and the provitamin $D_3$  is sensitive to UV light. However, molecules with a specified wavelength absorption can be engineered, as in the study [36] where they proved that by doping pure vanadia, they can change its optical absorption spectra as shown in Fig. 3.

For medical applications, the visible light is not dangerous to be used inside the human body. However, if shorter wavelengths are chosen to be used, as we proposed in our previous study, there are some standards that should be respected to insure the safety of its use. The UV light is not harmful in itself, but the radiation energy and time of exposure are what makes it harmful or not. Therefore, the American National Standards Institute (ANSI) defined the standard ANSI Z-136.1 [37], which gives the Maximum Permissible Exposure limits (MPEs) for users. The minimum MPE for the UV light is 3.0 $mJ/cm^2$ over 8h. The exposure limits given by the ANSI Z-136.1 are measured under worst-case conditions at the cornea of the human eye, because it is the most sensible part of our body to light. Therefore, if the generated UV energy and the exposure time are safe for the cornea, then they are safe to be used inside the body. While the predominant cause of UV radiation injury is a thermal process [38], a safe UV generator should be designed to generate non-thermal radiation by using pulsed beam instead of continuous one. The energy will be concentrated into a very short time, and this compression delivers the beam light more rapidly, which gives more power using less energy [39]. The shorter the time of the pulse, the less energy is used, which drastically decreases the risk of tissue burn. Thus, to safely generate UV light inside the body, the generator should use picosecond pulses or less, with energy less than 3 $mJ/cm^2$ . That can justify the chosen value of the pulse duration in our previous study , which is 4 picosecond pulse and 2 $mJ/cm^2$  as incident energy [40].

	\section{Receiver Observations}
The passive receiver observes and counts the number of molecules, which diffuses through it. In order to derive the expression of the expected molecules number at the receiver, we can use the differential equation of molecules diffusion described in Fick's second law (1). In this section, we use this differential equation as [24] to derive the expected molecules number for three different scenarios: a) the scenario without reaction, b) the scenario with enzyme reactions and c) the scenario with photolysis. The enzyme reactions can be modeled using the Michaelis-Menten kinetics presented as follows [41]: 

     \begin{equation}
     \ce{$E$ +$S$ <=>[\ce{$k_{-1}$}][\ce{$k_1$}]  $M_c$ ->[\ce{$k_2$}]$ E$ + $P$},
     \label{eq:3}
     \end{equation} Where $E$ is the concentration of enzymes, $S$ is the concentration of the substrate, $P$ is the concentration of the reaction product and $M_c$ is the complex enzyme-substrate called Michaelis complex. The rate constants of the reaction $k_n$ represent the speed at which each step of the reaction happens. The first reaction rate constant where n=1 represents the speed at which enzymes and substrates bind together. The second reaction rate constant (n=-1) represents the speed of their unbinding, and the last one (n=2) represents the substrate  degradation speed to the product. We can notice from (3) that enzymes are released intact after degradation and they can be reused and participate in future reactions. The photolysis reaction can be modeled as follows :
	
	\begin{equation}
	\ce{$S$ + $h$$\upsilon$ ->[\ce{$J$}] $P$},
	\label{eq:4}
	\end{equation} Where $h$ is Planck's constant, and $\upsilon$ is the photon frequency. The reaction rate $J$ is a very fast first order reaction decay constant, which degrades $S$ to $P$ by using the photon energy and we can write: 
	
	\begin{equation}
	\frac{\partial S}{\partial t} = -JS,
	\label{eq:5}
	\end{equation}
	
	The $J$ coefficient is dependent on solar radiation intensity and photo-physical properties of the substrate. We can calculate it as described in [42-Chap 9]:
	
	\begin{equation}
     J=\int_{\lambda} \phi(\lambda, T)\sigma(\lambda, T)F(\theta, \lambda) d\lambda,
     \label{eq:6}
  	 \end{equation} Where the function $ \phi$ is the quantum yield for the substrate as a function the wavelength $\lambda$ and temperature $T$. The function $\sigma$ is the absorption cross-section for a wavelength $\lambda$ and a temperature $T$, which can be gotten experimentally. $F (\theta, \lambda)$ is the light point irradiance (actinic flux) as a function of the wavelength $\lambda$ and the light zenith angle $\theta$.
  	
  	\subsection{Diffusion Without Degradation Reaction}
  	The received information at the receiver is represented by the observed impulse of molecules diffused from the transmitter at a distance $d$. The pdf expression of the expected molecules to be located inside the spherical receiver can be derived by solving the differential equation in (1). As in [24], we assume that the receiver is a point observer and that the concentration inside the receiver’s volume is uniform and equal to the expected concentration in the center. 
  	The solution of the differential equation can be written as:
  	
	\begin{equation}
		S(t) = \frac{NV}{8(\pi Dt)^{3/2}}\exp({\frac{-d^2}{4Dt}}),
		\label{eq:7}
	\end{equation} Where $S(t)$ is the probability density function (pdf) of the expected molecules number. $N$ is the number of molecules released by the transmitter for one bit. $V$ is the volume of the receiver, $D$ is the diffusion coefficient, $d$ is the distance between the transmitter and the receiver, $t$ is the time when molecules are detected by the receiver. The equation (7) is the starting point from which we begin the build of our proposed system model, and against which we compare and evaluate the two other scenarios.
	
  	\subsection{Diffusion With Enzyme Reaction}
  	In nature, enzymes are used to accelerate biochemical reactions to build or to degrade molecules. In most cases, enzymes are specific to only one reaction, because of their high selectivity. In the second scenario, enzymes are added in the medium and diffuse between the transmitter and the receiver. Binding to the molecules carrying the information, the enzymes influence the diffusion of the molecules, thus, the equation (1) becomes [24]:
  	
  	\begin{equation}
  	\frac{\partial Z}{\partial t} = D \nabla^2 Z + f(Z, t),
  	\label{eq:8}
  	\end{equation} Where $f(.)$ is the term of the reaction in (3) using the chemical kinetics principles, and $Z$ represents the different species of the reaction, molecules, enzymes, products and Michaelis complex. In this study, we are interested only in the term reaction of the molecules and the equation (8) can be written as: 	
	\begin{equation}
	\frac{\partial S}{\partial t} = D \nabla^2 S - k_1 S E + k_{-1} M_c,
	\label{eq:9}
	\end{equation} Where $k_1$ represent the rate of the binding between the molecules $S$ and the enzymes $E$, $k_{-1}$ is the rate of Michaelis complex ($M_c$) degradation into enzymes and molecules. 

	An analytical closed-form solution for the equation (9) is hard to find, and by making simplifying assumptions as explained in [24], we can derive a lower bound expression on the expected impulse response. To simplify the solution, we need to slow down the unbinding reaction by converging the rate $k_{-1}$  to zero, and accelerate the degradation reaction until we use all the enzymes in the medium, so that their concentration becomes constant $E_{tot}$ . Therefore, the lower bound expression can be written as:

	\begin{equation}
	S(t) \geqslant \frac{NV}{8(\pi Dt)^{3/2}}\exp({-k_1 E_{tot} - \frac{d^2}{4Dt}}),
	\label{eq:10}
	\end{equation} Where $E_{tot}$ is the total concentration of bound and free enzymes in the medium. The authors of the study in [24], proved that this lower bound expression can be more accurate and improves with time, and its accuracy is dependent on the accuracy of the two assumptions made about the reaction. We can notice that the equation (10) can directly be compared to the equation (7). We see that the addition of enzymes in the medium increases the decaying exponential caused by the distance. If we increase the binding rate or the total concentration of enzymes, we can increase the decaying speed, which decreases the ISI. However, that will also decrease the strength of the useful signal, because we cannot control the time of enzyme reactions. Although, this can be possible with photolysis reactions.
	
		\subsection{Diffusion With Photolysis Reaction}
	The third scenario is similar to the second one, but instead of degrading the molecules with enzymes, we use the energy of photons. The diffusion of molecules in this case follows two phases, a phase without light and a phase with light. The diffusion in the first phase is the same as in the first scenario (without any reaction), and it can be described with equation (1). The diffusion in the second phase when the light is turned ON changes, because when the photon hits the molecule, the energy that degrades it also slightly change its direction. Light influence the diffusion of molecules as enzymes do in the second scenario, and the second phase can be described with the equation (8). Therefore, the differential equation in this scenario is the sum of (1) and (8), and it can be written as:
	
	\begin{equation}
	\frac{\partial S}{\partial t} =
	\begin{cases}
	D \nabla^2 S, &  t<T_{op}, \\
	D \nabla^2 S + f(S, t), &  t \geqslant T_{op}.
	\end{cases}
	\end{equation}
	
	Where $S$ is the concentration of the released molecules, $T_{op}$ is the optimal time to turn the light ON, which will be derived in the next section.  $f(S, t)$ is the term of the reaction in (4), and by using (5) we can write (11) as:
	
    \begin{equation}
	\frac{\partial S}{\partial t} =
	\begin{cases}
	D \nabla^2 S, &  t<T_{op}, \\
	S(D \nabla^2 - J), &  t \geqslant T_{op}.
	\end{cases}
	\end{equation} Where $J$ is the reaction rate coefficient. 
	
	The assumptions made to simplify the solution of the equation (9) in the case of enzymes do not exist in the case of photolysis. However, we assume that the intensity of light is uniform around the receiver with a radius $r$ = $d$ so that the degradation of molecules will be homogeneous, which is not the case in the real system. Under the boundary conditions of the third scenario, the solution of the equation (12) gives a lower bound expression on the expected number of molecules at the receiver. Then, the expected impulse response of the photolysis scenario can be written as:
	
	\begin{equation}
	\frac{\partial S}{\partial t}  \geqslant
	\begin{cases}
	\frac{NV}{8(\pi Dt)^{3/2}}\exp({\frac{-d^2}{4Dt}}), &  t<T_{op}, \\
	\frac{NV}{8(\pi D T_{op})^{3/2}}\exp({-Jt - \frac{d^2}{4D T_{op}}}), &  t \geqslant T_{op}.
	\end{cases}
	\end{equation}
	
	The equation (13) shows that the emitting light acts as an additional decaying exponential, which degrades the molecules and mitigates ISI. Before emitting the light, the molecules diffuse in the medium without being degraded, which gives time to a maximum number of molecules to be observed by the receiver, so that the strength of the useful signal is maintained. After emitting the light in an optimal time, the additional decaying exponential degrades the remaining molecules, which mitigates the ISI. Increasing the coefficient $J$ will result in a faster decaying, which can be done by increasing the light intensity, increasing the energy of the photons or using a wavelength for which the absorption of molecules is in its maximum. The time of light emission is important, as shown in our previous study. In the next section, we present the derived expression of light emission’s optimal time $T_{op}$, and the expression of the bit error probability.
	
	\section{Performance Evaluation }	
	In order to evaluate the performance of the proposed system and compare it with the other two scenarios, we simulate each scenario using AcCoRD simulator [43]. We also use the  Interference-to-Total-Received molecules metric (ITR) and the derived bit error probability. The Actor-based Communication via Reaction-Diffusion (AcCoRD) contains a microscopic and mesoscopic hybrid algorithm, which increases its computational efficiency and its simulation accuracy. 
	
	\subsection{Light Emission's Optimal Time}
	
	The time of light emission at the receiver is crucial for the quality of the received signal. If the light is emitted too early, the strength of the signal will be decreased, and if it is emitted too late, the ISI mitigation will be decreased. Therefore, in order to have an optimal impulse response at the receiver, an expression of the optimal time for light emission must be derived. First, we simulate the proposed system with different light emission times, and with a fixed distance between the transmitter and the receiver (5 $\mu m$). The results of the simulation showed that the optimal time to emit the light is when the expected number of molecules at the receiver is in its maximum. The optimal time in the simulation result is $t=0.04s$ as shown in Fig. 4. Since this maximum only occurs in the first phase of our proposed system, which means when the model follows the first case of this study, the expression of the optimal time should be derived from (7). A simple calculation of the equation’s derivative with respect to $t$ is needed to find the time of its peak and thus, the expression of the optimal time to emit light can be written as follows:
	
	\begin{equation}
	T_{op} = \frac{d^2}{6D},
	\label{eq:14}
	\end{equation} 
	
	 The optimal time is proportional to the distance between the transmitter and the receiver, which implies that the proposed system would favorably use a fixed transmitter and a fixed receiver. However, if the distance $d$ of the system is variable, the receiver should have a mechanism that allows it to know the position of the transmitter before emitting the light. The variable distance of the system is an interesting problem, which we will leave for future work.
	
	\subsection{Interference-to-Total-Received molecules (ITR)}
	
	We use the ITR in this study to evaluate the performance of the proposed system model, because this metric clearly highlights the enhancement of the signal strength and the ISI mitigation at the same time. The ITR metric analysis provides an in-depth understanding of the difference between the three studied scenarios and it can be defined as [25]:
		
	\begin{figure}
	\centering
	\includegraphics[width=0.975\linewidth]{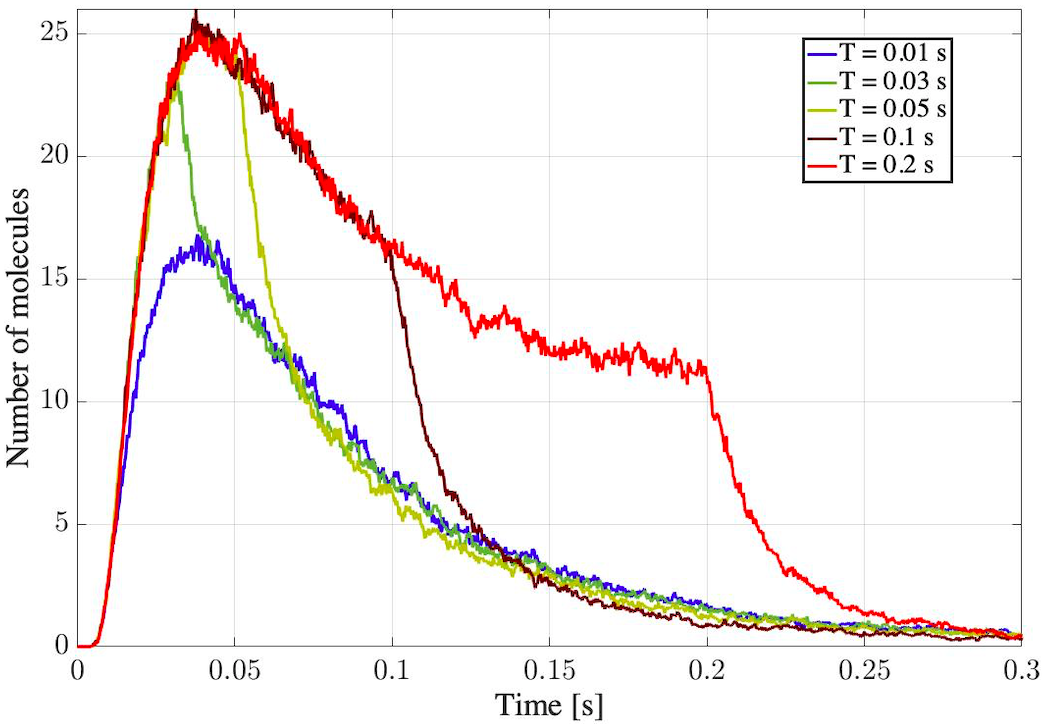}
	\caption{Effect of the light emission's time variation on the impulse response.}
	\label{fig:4}
    \end{figure} 

	\begin{equation}
    ITR(t_s, t_{end}) = \frac{F(t_{end}))- F(t_s)}{F(t_{end})}.
    \label{eq:15}
    \end{equation} 	Where $F(t_{end})$ is  the total expected number of molecules at the receiver, and $F(t_s)$ is the expected number of molecules at symbol period $t_s$, which we take it in this study as in [27] ($t_s = 0.1s$).
    
    The ITR calculates the normalized difference between the total expected molecules and the molecules observed before the symbol period. This difference represents the ISI problem. If the calculated difference is small, that means the received signal is strong, and the system is better at ISI mitigation. Therefore, in our study, the smaller the ITR, the better the performance of the system.
	
	\subsection{Bit Error Probability}
	
	In this study, we modulated the signal with a binary shift in the number of molecules. We assume that the receiver contains a simple detector, which compares the received molecules number with a decision threshold value $\zeta$. The threshold can determine whether the send bit is 0 or 1 and thus, we define the error in our study as the decided bit value that is not equal to the actual bit value sent from the transmitter. In this paper, we considered only the case of transmitting one bit, because the purpose of the study is to prove that the proposed system can mitigate ISI without decreasing the strength of the signal, and one bit is sufficient to do it. The case of bits sequence will be considered in future work. 
	Let's consider $\beta$ the information bit sent by the transmitter, which it can be either 0 or 1. The $a$ $priori$ probabilities of the bit are $P_0$ and $P_1$=1- $P_0$, which are the probabilities of $\beta$ to be 0 or 1 respectively. Let's consider $\widehat{\beta}$ the received bit at the receiver at time $T_{op}$, $\widehat{\beta}$ should be equal to $\beta$  otherwise, there will be an error. The detection scheme can be written as follows:
	
	\begin{equation}
	\text{\footnotesize $\widehat{\beta} $}=
	\begin{cases}
 	\text{\footnotesize $1,  S(T_{op}) \geqslant \zeta,$} \\
    \text{\footnotesize	$0,  S(T_{op})< \zeta.$}
	\end{cases}
	\end{equation} 
	
	The threshold $\zeta$ is chosen empirically by trying several values and then choosing the one with the smallest probability of error. Since we're considering the case of only one bit, $P_0$ is not the best option to study the error probability, because there would be no molecules to be observed at the receiver, so instead, we study the probability $P_1$. The probability of $\beta$=1 is the probability of $S(T_{op})$ to be bigger or equal to the threshold value which can be written as $P_r (S(T_{op})\geqslant \zeta)$. If we assume that each molecule is independent of all the other ones, and if we put $N$=1, we can use the lower bound expression on the expected molecules number in equation (13) as the probability density function of a single molecule’s location at any time. As we mentioned before, we assume that the receiver is a point observer and the concentration of the molecules within its volume is uniform. In order to lighten the writing of the equation (13), we note the first phase of the equation as $W_R$ for (without reaction). Thus, the lower bound expression on the probability $P_S(t)$ that one molecule shortly exposed to the light is observed at the receiver can be written as:
	
	\begin{equation}
	P_S(t) =
	\begin{cases}
	W_R, &  t<T_{op}, \\
	W_R \exp(-Jt), &  t \geqslant T_{op}.
	\end{cases}
	\end{equation}
	
	The equation (17) can be met with equality if we assume that the light intensity is uniform and the molecules degradation is homogenous. So the probability that all molecules shortly exposed to the light are observed at the receiver is $N$ times the probability of one molecule. The molecules have two possibilities in the medium, whether they pass through the receiver and being observed or they miss the receiver without being observed. Thus, the probability $P_r (S(t)\geqslant \zeta)$ follows a binomial distribution [44] and can be written as:
	
	\begin{equation}
	P_r (S(t)\geqslant \zeta) = \sum_{q=\zeta}^{N} \binom{N}{q} P_S(t)^q(1-P_S(t))^{N-q},
	\end{equation}
	
	By substituting (17) in the binomial equation (18), we write:
	
	\begin{equation}
	\text{\footnotesize $P_r (S(t)\geqslant \zeta)$}=
	\begin{cases}
	 \sum_{q=\zeta}^{N} \binom{N}{q} W_R^q(1-W_R)^{N-q}, t<T_{op}, \\
	 \sum_{q=\zeta}^{N} \binom{N}{q} (W_R\exp(-Jt))^q  \\ \quad (1-(W_R\exp(-Jt)))^{N-q},  t \geqslant T_{op}.
	\end{cases}
	\end{equation}
	
	As we mentioned above, the error probability in this study is higher if $P_r (S(t)< \zeta)$ when the sent bit is 1 or if $P_r (S(t)\geqslant \zeta)$ when the sent bit is 0. In our case of a single bit transmission, the error probability is:
	
	\begin{equation}
	P_e = P_1(P_r (S(t)< \zeta)),
	\label{eq:20}
	\end{equation} Where $P_1$ is the $a$ $priori$ probability of the bit to be 1, and we put it as $P_1$=0.5. (20) can also be written as:
	
	\begin{equation}
	P_e = P_1(1 - P_r (S(t)\geqslant \zeta)),
	\label{eq:21}
	\end{equation} 
	
	However, the equation in (19) can give exact results only for one molecule probability, and the results become very difficult to evaluate for greater values of $N$, as in molecular communication case. Poisson distribution is a special case of the binomial, where the number of trails is large (which is the total number of molecules in our case) and the probability of success in each trial is small (the probability of one molecule to be observed). So, we can approximate the binomial distribution in equation (19) for larger values of $N$ and small values of $P_S (t)$ to a Poisson distribution with mean and variance $S(t)=NP_S (t)$. Therefore, the probability $P_r (S(t)\geqslant \zeta)$ can be written as [24]:
	
	\begin{equation}
	P_r (S(t)\geqslant \zeta)|_{Poiss} = 1- \exp(-S(t)) \sum_{q=0}^{\zeta - 1} \frac {S(t)^q}{q!},
	\end{equation}
	
	By substituting (17) in the Poisson equation (22), we write:
	
	\begin{equation}
    \text{\footnotesize $P_r (S(t)\geqslant \zeta)|_{Poiss}$}=
	\begin{cases}
	1- \exp(-W_R) \sum_{q=0}^{\zeta - 1} \frac {W_R^q}{q!},   t<T_{op}, \\
	1- \exp(-W_R\exp(-Jt)) \\  \quad \sum_{q=0}^{\zeta - 1} \frac {(W_R\exp(-Jt))^q}{q!},  \quad \;  t \geqslant T_{op}.
	\end{cases}
	\end{equation}
	
	However, the factorial in the Poisson distribution increases the computational burden. By using the Central Limit Theorem (CLT) [44], it is generally considered appropriate to approximate the Poisson by a Gaussian distribution when the mean of the Poisson is bigger than 20. Despite the fact that Poisson gives more accurate approximation, in this study, we consider the Gaussian distribution because of its computational efficiency, by assuming that $P_S(t)$ is not close to zero or one. The study in [24] proved that Gaussian approximation gives an acceptable loss in accuracy, and we validated this result in this study as we will see in the next section. The mean and variance of the approximated Gaussian are $S(t)$ and $S(t)(1-P_S(t))$ respectively, and we write:
	
	\begin{equation}
	P_r (S(t)\geqslant \zeta)|_{Gauss} = \int_{q=0}^{\zeta - 1} \frac{ \frac {-(q-S(t))^2}{\exp(2[ \,S(t)(1-(P_S(t)))] \,^2} }{\sqrt{2\pi S(t)(1-(P_S(t)))}},
	\label{eq:24}
	\end{equation}
	
	The solution of this integral cannot be expressed in terms of elementary functions. However, by using the error function, it can be approximated as [45]:
	
	\begin{equation}
	\text{\footnotesize $P_r (S(t)\geqslant \zeta)|_{Gauss} =   \frac{1}{2} \Bigg [1+ \erf \Big (\frac{\zeta - S(t)}{\sqrt{2S(t)(1-P_S(t)) }} \Big ) \Bigg ],$}
	\label{eq:25}
	\end{equation}

By substituting (17) in the error function (25), and putting the variances $W_R(1-\frac{W_R}{N})$ and $W_R\exp(-Jt)(1-(\frac{W_R}{N}\exp(-Jt)$ as $\rho_1$ and $\rho_2$ respectively, we write:

\begin{equation}
\text{\footnotesize $P_r (S(t)\geqslant \zeta)$}=
\begin{cases}
\frac{1}{2} \Big [1+ \erf \Big (\frac{\zeta - W_R}{\sqrt{2\rho_1}} \Big ) \Big ], & t<T_{op}, \\
\frac{1}{2} \Big [1+ \erf \Big (\frac{\zeta - (W_R\exp(-Jt))}{\sqrt{2\rho_2}}  \Big ) \Big ],& t \geqslant T_{op}.
\end{cases}
\end{equation}

To evaluate the error probability $P_e$ of the proposed system, we can calculate the value $P_r (S(t)\geqslant \zeta)$ in equation (21) by using either the equation (19), (23) or (26). In this study, we favored the computational efficiency over the accuracy by using the equation (26) in the bit error probability calculation.

	 \section{Numerical and Simulation Results} 
	
	The environment parameters for each simulated scenario in this study are chosen so that the simulation gives clear results without increasing the calculation burden. The distance between the transmitter and the receiver as an example is chosen to be between 5 and 10 $\mu m$ because in that range, the simulation gives clear results with the used number of molecules $N$=10,000. After 10 $\mu m$, the impulse response drastically decreases so that the graph becomes hard to read. As for the biochemical parameters, we choose them based on biological studies and other studies in the literature, and we decreased the reaction rates to simulate the worst case in the last two scenarios, as shown in table 1.

	In order to simulate the photolysis reaction, we use fast virtual enzymes in the medium to play the role of the light degradation. The choice of virtual enzymes to simulate the first  order degradation rate of the photolysis is a normalization attempt to facilitate the comparison between our proposed method that uses light and the method that uses enzymes. In this study, we simulate the worst case in the three scenarios to use them as upper bound results. The worst case in the photolysis reaction is when the viscosity of the medium is high, and light intensity is reduced in far distances from the receiver. To take into account this light intensity reduction in our simulation, we add virtual spheres with increasing diameters around the receiver, and we fill each sphere with fast virtual enzymes. The concentration of the enzymes is degraded with the increase in the spheres diameter to capture the influence of light intensity reduction on the proposed system.

	The simulation results presented in this paper are the average of the received molecules number in each iteration, with a maximum number mean of 24.78 $\pm$ 0.49 and 19 as a standard deviation. The confidence interval of the results is 99$\%$ with a margin error of 0.49. The sample mean is between [24.29-25.27].

		\subsection{Impulse Response}
	The impulse responses at the receiver for the three studied scenarios are presented in Fig. 5. We can notice that in the first scenario (without reaction), the ISI is very high, presented with a heavy tail, which is drastically decreased in 
		\begin{table}[h]
		\caption{System Parameters Used for Simulation.}
		\centering{
			\begin{tabular}{|l|c|r|r|}
				
				\hline
				Parameters & Scenario 1 & Scenario 2 & Scenario 3 \\
				\hline
				Released molecules &10,000&10,000 & 10,000 \\ \hline
				Diffusion coefficient ($m^2/s$) & $10^{-10}$ & $10^{-10}$ & $10^{-10}$ \\ \hline
				Simulation repetition  & 100 & 100 & 100  \\ \hline
				Simulation time ($s)$  & 0.5  & 0.5  &  0.5  \\ \hline
				Distance ($\mu m$)  & 5  & 5 & 5 $\sim$ 10  \\ \hline
				Receiver radius ($\mu m$)  & 5 & 5 & 5  \\ \hline
				Number of enzymes & - & 12980 & 12980   \\ \hline
				Binding rate ($ns^{-1}$ ) & - & 5 & 10  \\ \hline
				Unbinding rate ($s^{-1}$ ) & - & 1 &  10  \\ \hline
				Degradation rate ($s^{-1}$ ) & - & 1  & 10   \\ \hline
				Symbol period $t_s$ ($s$)   & 0.1  & 0.1  & 0.1  \\ \hline
				Simulation time step $\Delta t$ ($\mu s$) & 0.2 & 0.2 & 0.2   \\ \hline
				
			\end{tabular}
		}
		\label{tab:tab1}
	\end{table} the second scenario with the enzyme presence. However, the enzyme also decreases the amplitude of the signal with 36$\%$, which is considered in the literature as a trade-off, a strong signal creates more ISI, and mitigating ISI weaken the signal. Nonetheless, in the case of the photolysis, a clear comparison between the three scenarios shows that the proposed system mitigates ISI faster, without decreasing the strength of the signal, because of the switching ability of the light. The ISI in the first scenario is in its worst case because the simulated environment is bounded, and the receiver is close to the reflected molecules in its boundaries, which increases the ISI possibility.

		\begin{figure}
		\centering
		\includegraphics[width=\linewidth]{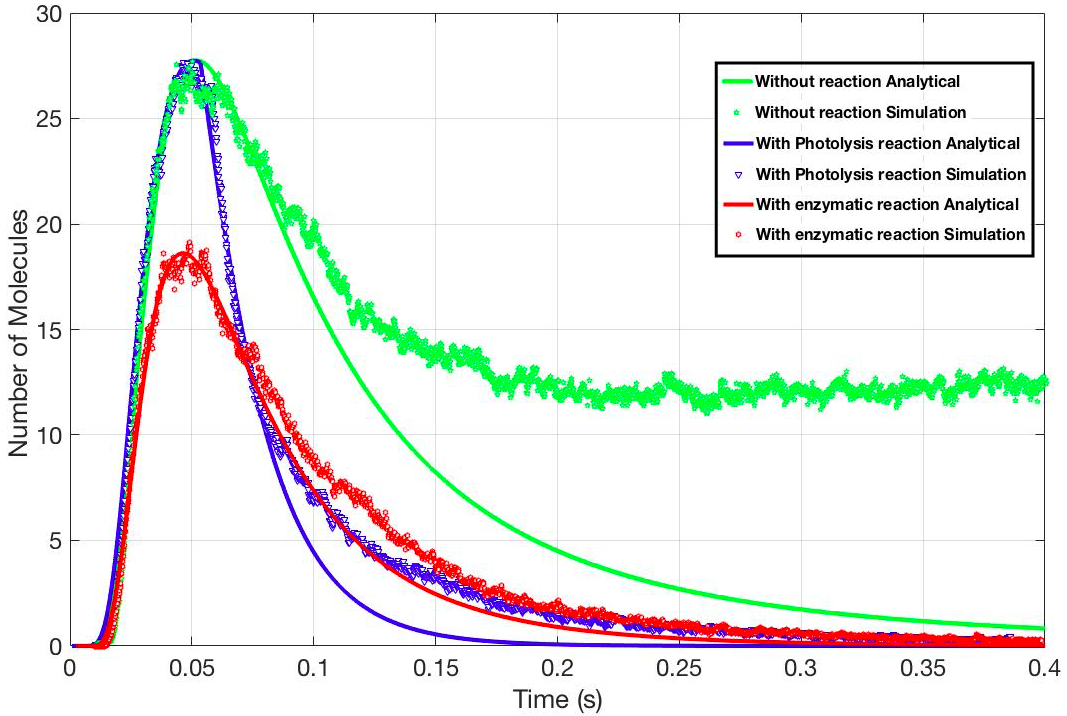}
		\caption{Analytical and simulation results of the impulse response for the three studied scenarios.}
		\label{fig:5}
    	\end{figure} 
    
    	\begin{figure}
    	\centering
    	\includegraphics[width=\linewidth]{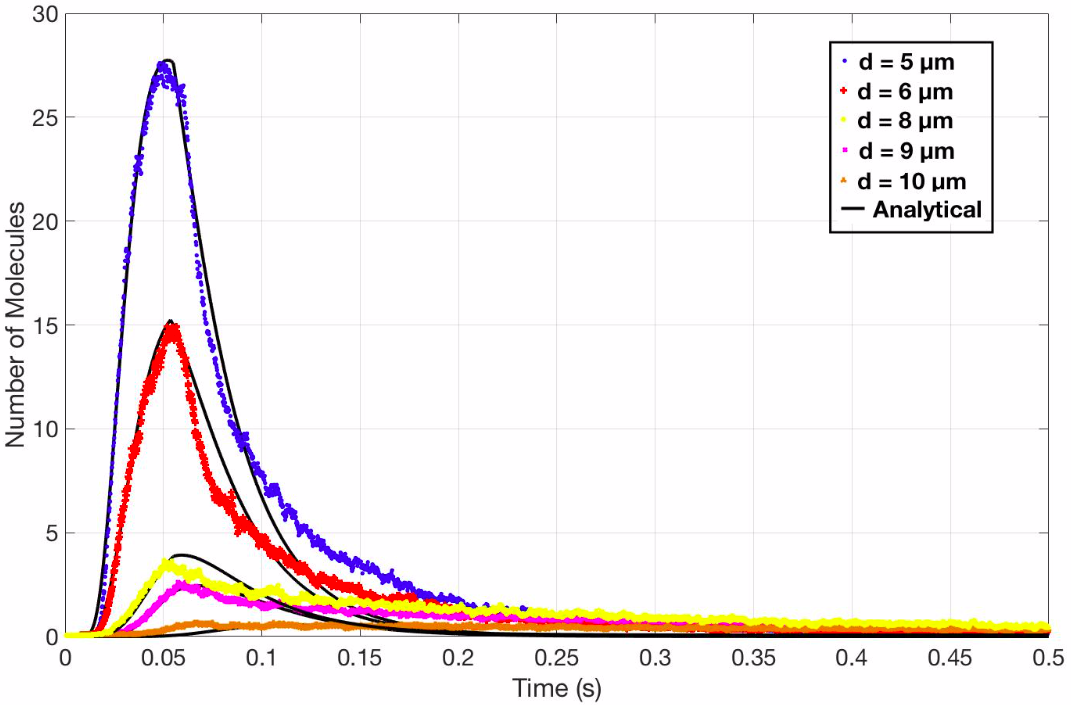}
    	\caption{Distance influence on the impulse response of the proposed system.}
    	\label{fig:6}
        \end{figure} 
	
	The Fig. 5 also shows a comparison between the simulation results and the lower bound expressions discussed in section 3. We can notice that the analytical results are accurate for the amplitude of the signal, but not accurate for ISI mitigation, due to the fact that the simulations represent the worst case and the lower bound expressions represent the best case for the ISI. The actual response at the receiver will be between the analytical and the simulation results. It’s clear though that ISI mitigation of the worst case in our proposed system is comparable to the best case of the enzymes and without decreasing the signal strength.

	The Fig. 6 shows the influence of the distance on the impulse response of the system. The distance is inversely proportional to the expected number of molecules as shown in the equation (13) of the proposed system model. We can notice that the analytical results accurately predict the maximum number of molecules for distances between 5 and 10 $\mu m$.
	
		\subsection{ITR Evaluation}	
	The ITR defined in (15) is a good metric to evaluate the performance of the proposed system, because it clearly highlights the enhancement in the signal strength and ISI mitigation at the same time. In this study, the smaller the ITR, the better the performance of the system as explained in the previous section.

	The Fig. 7 shows the ITR values of the three studied scenarios and as expected, the photolysis of our proposed system has the smallest ITR value. We can see that the first scenario where there is no degradation reaction has better ITR value than the scenario of enzymes. The bigger value of ITR in the presence of enzymes can be explained by the decrease in the signal strength, which drastically increases the ITR. The evaluation of the three studied scenarios using ITR validates the numerical and the simulation results, and proves that the performance of the proposed system is better in mitigating ISI, without decreasing the useful signal.

		\begin{figure}
		\centering
		\includegraphics[width=\linewidth]{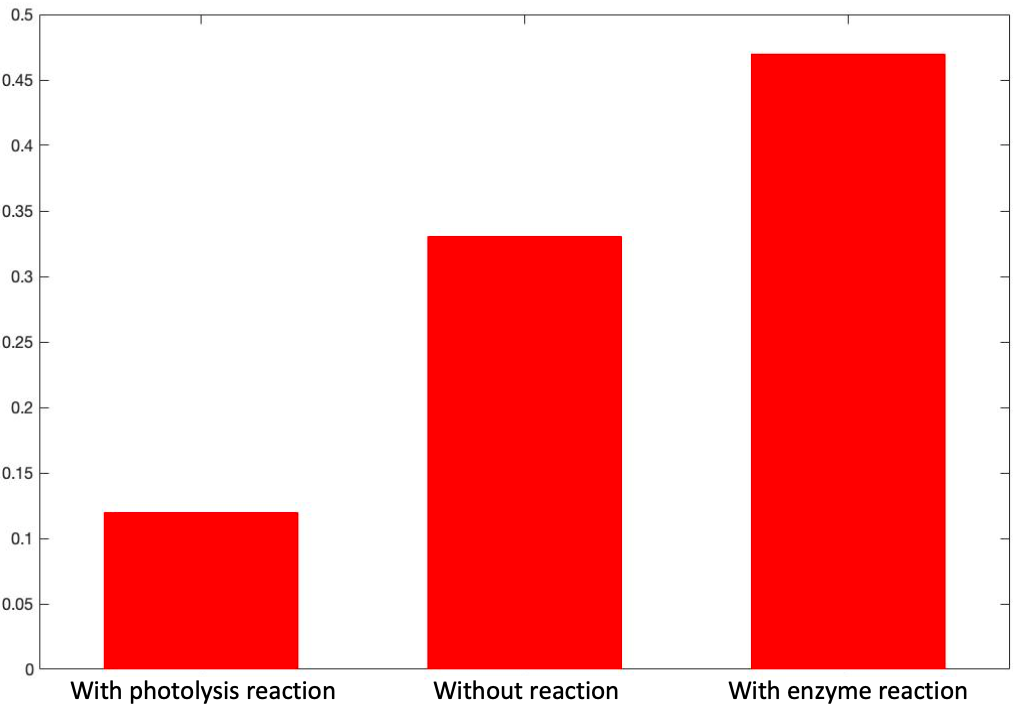}
		\caption{ITR values of the three studied scenarios, 1) without reaction, 2) with enzymes, 3) with photolysis.}
		\label{fig:7}
	   \end{figure} 
	 
	 	\subsection{Bit Error Probability Evaluation}	
	 		In the previous section, we derived the Poisson and Gaussian expressions to approximate the probability of molecules number observed at the receiver. Fig. 8 shows that Poisson approximation follows the simulation more accurately than Gaussian, which has some loss in accuracy as proved in [24]. However, the Gaussian has more computational efficiency and its accuracy loss is not very large, and it still can be used to predict the probability of error. We can also notice that the photolysis scenario has a little bit less probability of errors compared to the enzyme scenario for a threshold value between 3.5 and 5. Nonetheless, if we increase the threshold value $\zeta$ $\geqslant$ 5, we can note a considerable difference between the proposed system using photolysis and the enzyme system in terms of error probability. In Fig. 9, we can clearly see that the performance of the proposed scenario is better than the other two scenarios. With a threshold value $\zeta$=7, the proposed system can have less than 10$^{-3}$ probability to make errors. For small threshold values, enzymes and photolysis have similar performance, which is logical because the maximum number of molecules expected at the receiver is bigger than the threshold value. The considerable difference between the two systems for higher threshold values can be explained with the significant decrease in the signal strength when using enzymes (36$\%$ loss). When the threshold value becomes bigger than the maximum number of molecules, the receiver has more probability to make errors, and thus, the bigger the amplitude of the signal, the better its performance. The scenario that uses no reaction has the worst error probability with 10$^{-1}$, because of the heavy tail that causes the ISI.

	\begin{figure}
	 		
	\centering
	 		
	\includegraphics[width=\linewidth]{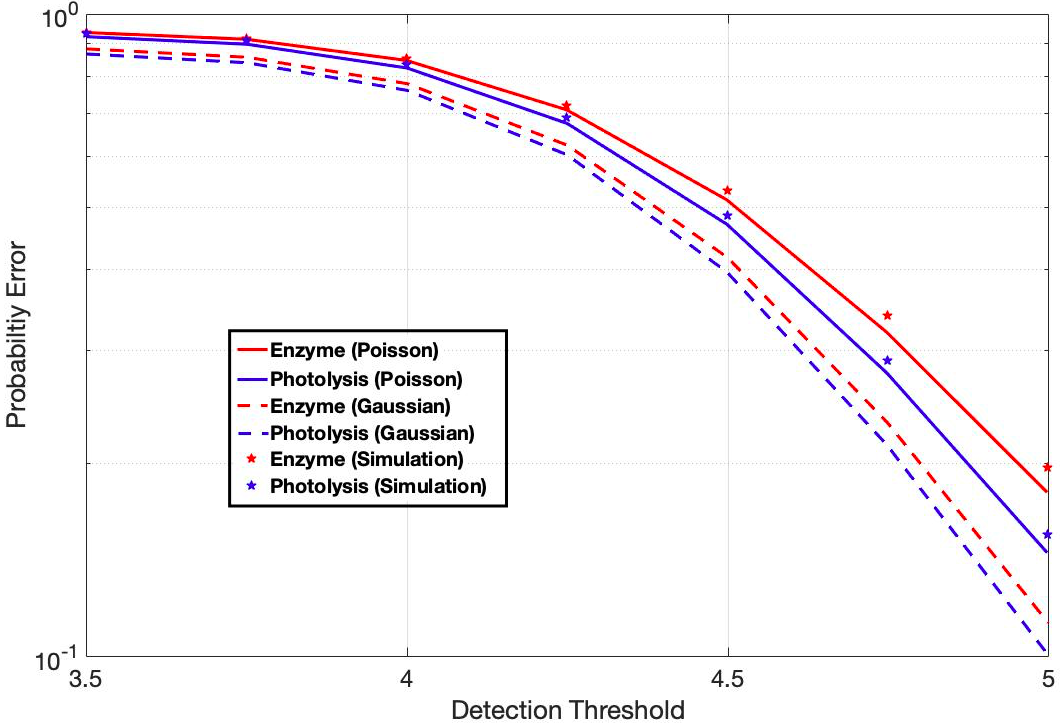}
	 		
	\caption{Accuracy of Poisson and Gaussian approximations for the enzyme and the photolysis reactions.}
	 		
	\label{fig:8}
	 		   
\end{figure}

\begin{figure}
	 		
	\centering
	 		
	\includegraphics[width=\linewidth]{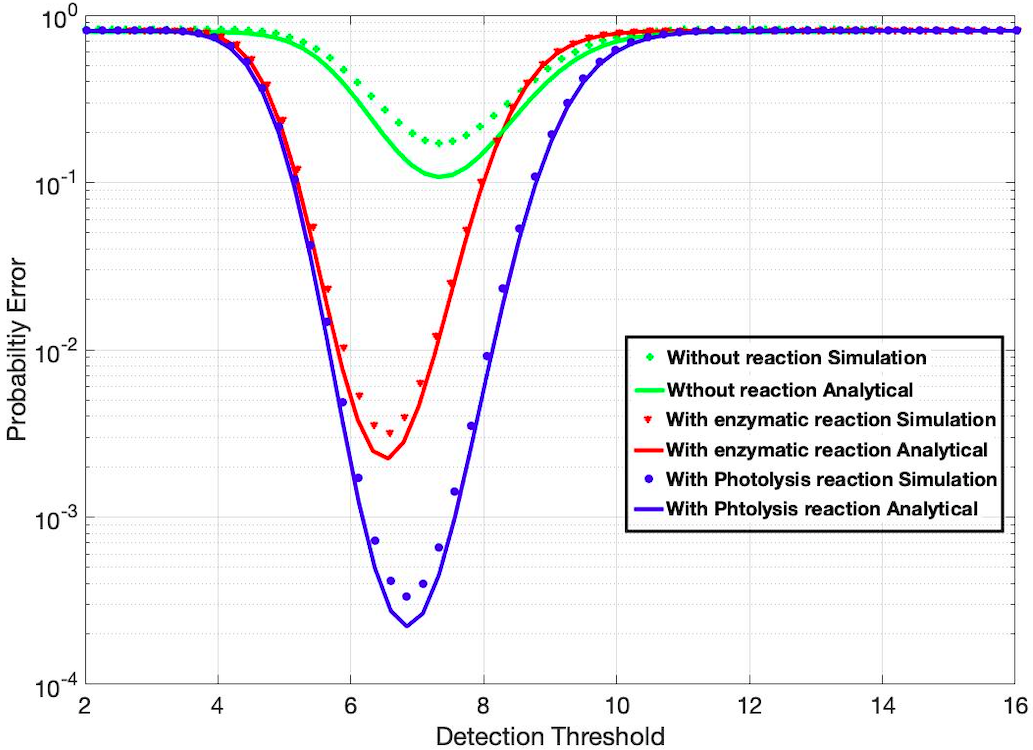}
	 		
	\caption{Bit error probability of the three studied scenarios as a function with the detection threshold.}
	 		
	\label{fig:9}
	 		
   \end{figure}

	 		We can also notice in Fig. 9 that the analytical probability of error for the three studied scenarios is smaller than the simulation when it reaches the threshold values that make the minimum error. As explained in the previous section, the analytical expressions are lower bounds on the systems actual response, and the simulations are used as upper bounds. Therefore, the actual value of the error probability is expected to be between the analytical and the simulation results, and again, the worst case of our proposed system outperforms the best case of the two others. We can conclude that using photolysis reactions can mitigate ISI faster, without decreasing the useful signal, which clearly enhance the performance of molecular communication.

	 \section{Conclusion}
	Diffusion-based molecular communication is one of the most promising bio-inspired paradigms for the communication at nano scale. A lot of work was proposed in literature to mitigate its main challenge, which is ISI. In this paper, we extended our proposed system in [27], which uses photolysis reaction instead of enzymes to degrade the remaining molecules in the medium. Unlike the enzymes, the switching ability of light allows the receiver to mitigate ISI faster without decreasing the strength of the signal. We presented a comparison study between three scenarios, a system without reaction, with enzymes and with photolysis. We then derived a lower bound expression on the expected number of molecules to be observed at the receiver, while shortly exposed to light. We also derived the bit error probability and used it alone with interference-to-total received molecules metric to evaluate the performance of the proposed system. The analytical results validated with simulation showed a visible enhancement when using photolysis reaction, which improves the performance of the diffusion-based molecular communication.

	Our future work will include considering the variable distance problem discussed in the section 4, the case of bits sequence and the case of absorbing and reflective receiver. We will also consider an experimental study at the macro scale using UV light to degrade molecules, and then we will provide a comparison study between analytical, simulation and experimental results.

\end{document}